\documentclass[review]{elsarticle}

\usepackage[colorlinks]{hyperref}
\usepackage{amsmath}
\usepackage{amssymb}
\usepackage{todonotes}
\usepackage{multirow}
\usepackage{lineno,hyperref}
\usepackage{url}
\usepackage{soul}
\usepackage{subfigure}
\usepackage{tabularx}
\usepackage{caption} 
\makeatletter
\def\ps@pprintTitle{%
\let\@oddhead\@empty
\let\@evenhead\@empty
\def\@oddfoot{}%
\let\@evenfoot\@oddfoot}
\makeatother

\journal{}
\begin{document}

\begin{frontmatter}

\title{Evaluation of SNOLAB background mitigation procedures through the use of an ICP-MS based dust monitoring methodology}


%\author[mymainaddress,mysecondaryaddress]{Author list}

%\author[mysecondaryaddress]{Corresponding author\corref{mycorrespondingauthor}}
%\ead{X@X.gov}

%\address[mymainaddress]{Affiliation 1}
%\address[mysecondaryaddress]{Affiliation 2}

\author[pnnl]{M.L. di Vacri\corref{mycorrespondingauthor}}
\author[snolab,lu]{S. Scorza\fnref{myfootnote}}
\author[pnnl]{A. French} 
\author[pnnl]{N.D. Rocco} 
\author[pnnl]{T.D. Schlieder}
\author[pnnl]{I.J. Arnquist}
\author[pnnl]{E.W. Hoppe}
\author[snolab,lu]{J. Hall} 
%\author[pnnl]{M.L. di Vacri}

\address[pnnl]{Pacific Northwest National Laboratory, Richland, WA 99352, USA}
\address[snolab]{SNOLAB, Lively, ON P3Y 1N2, Canada}
\address[lu]{Laurentian University, Department of Physics, Sudbury, ON P3E 2C6, Canada}
%\address[lpsc]{Univ. Grenoble Alpes, CNRS, Grenoble INP, LPSC-IN2P3, 38000 Grenoble, France}
\cortext[mycorrespondingauthor]{marialaura.divacri@pnnl.gov}
\fntext[myfootnote]{Currently at Univ. Grenoble Alpes, CNRS, Grenoble INP, LPSC-IN2P3, 38000 Grenoble, France}

\begin{abstract}

Dust particulate fallout on materials in use for rare-event searches is a concerning source of radioactive backgrounds due to the presence of naturally occurring radionuclides $^{40}$K, $^{232}$Th, $^{238}$U, and their progeny in dust. 
Much effort is dedicated to inform radioactive backgrounds from dust and evaluate the efficacy of mitigation procedures. A great portion of such effort relies on fallout models and assumed dust composition. In this work, an ICP-MS based methodology was employed for a direct determination of fallout rates of radionuclides and stable isotopes of interest from dust particulate at the SNOLAB facility. Hosted in an active mine, the SNOLAB underground laboratory strives to maintain experimental areas at class 2000 cleanroom level. This work validates the mitigation procedures in place at SNOLAB, and informs dust backgrounds during laboratory activities. Fallout rates of major constituent of the local rock were measured two to three orders of magnitude lower in the clean experimental areas compared to non-clean transition areas from the mine to the laboratory. A $ca.$ two order of magnitude increase in stable Pb fallout rate was determined in an experimental area during activities involving handling of Pb bricks. Increased $^{40}$K, $^{232}$Th, and $^{238}$U fallout rates were measured in clean experimental areas during activities generating particulate.

\end{abstract}

\begin{keyword}
%\texttt{Word 1}
Surface contamination \sep Dust radioactive background \sep Background mitigation \sep ICP-MS \sep Ultra-low background materials \sep Cleanroom dust
%\MSC[2010] 00-01\sep  99-00
\end{keyword}

\end{frontmatter}

%\linenumbers

\section{Introduction}\label{intro}

Detectors studying rare events, such as neutrinoless double beta decay or dark matter, require construction and operation in rigorously ultra-low background (ULB) conditions in order to attain target sensitivities. Deep underground facilities typically provide ideal shielding from cosmic radiation \cite{paoluzi1989gran, bettini2012canfranc, smith2012snolab, lesko2015sanford, murphy2012boulby}, and shielding materials around the detectors are used to attenuate backgrounds from soil and rock. 

Radiopurity requirements for construction materials are extremely stringent, typically in the $\mu$Bq$\cdot$kg$^{-1}$ range, and can be even more stringent the closer the material is to the active target or the larger the mass is incorporated into the detector. Impurities of naturally occurring radionuclides $^{40}$K, $^{232}$Th, $^{238}$U, and their progeny in materials generate backgrounds. Sourcing and screening materials for ULB detectors is challenging yet imperative, since material backgrounds are a limit to detector sensitivities. Extensive assay campaigns are performed in order to source the most radiopure materials \cite{alimonti1998ultra, abgrall2014majorana, aprile2011material, aznar2013assessment, leonard2008systematic, budjavs2009gamma, grate2016icp, arnquist2019mass, akerib2020lux}. In order to reduce the impact of particle deposition on surfaces, ULB detectors are assembled and operated in cleanroom facilities, where the air particulate is significantly reduced compared to regular environments, with the level of particulate reduction depending on the characteristic of the specific facility \cite{Cleanair}. Even so, fallout of dust particulate on ULB detector materials remains a concerning source of radioactive backgrounds, given the presence of radionuclides $^{40}$K, $^{232}$Th, $^{238}$U, and their progeny in dust. The chemical composition of dust generally reflects the local composition of soil, and can be affected by anthropogenic activities. 
The level of natural K in soils is typically on the order  1$\%$, corresponding to 1 part per million, or ppm, of $^{40}$K, based on potassium natural isotopic composition. The content of $^{232}$Th and $^{238}$U is on the order of a few ppm. As a result, the soil radioactivity level is in the Bq$\cdot$kg$^{-1}$ range, several orders of magnitude greater than radiopurity requirements of ULB detectors. Surface contamination from dust fallout, as well as material handling, storage, transportation, and assembly can therefore significantly increase ULB material backgrounds. Understanding such backgrounds and contamination mechanisms is necessary to develop successful mitigation techniques and reliable handling procedures. 

Within the low-background community, much effort is being dedicated to understanding and mitigating backgrounds from dust particulate fallout. Dust monitoring techniques include the use of witness plates to collect dust in relevant locations, followed by X-Ray Fluorescence (XRF) analysis for Th and U proxy elements, e.g., Ca and Fe \cite{boger2000sudbury, SNO, SNO2, SNO3}, or optical and fluorescence microscopy analysis \cite{akerib2020lux}. It is worth noting that dust background predictions are oftentimes performed based on models and assumed dust chemical composition. In previous work \cite{di2021direct}, we have developed a method for the direct determination of fallout rates of contaminants of interest from dust particulate deposition on ULB materials. The method includes collection of dust on ULB media (for convenience, we have used ULB vials), followed by particulate dissolution and ultra-sensitive inductively coupled plasma mass spectrometric (ICP-MS) analysis. The study demonstrated that dust composition in cleanrooms is strongly affected by ongoing activities, and that indirect assessment methods that rely on assumed elemental compositions of dust may result in inaccurate background predictions.

In this work, we have employed the method described in \cite{di2021direct} to monitor key experimental areas at the SNOLAB facility during standard day-to-day activities and in a specific area where a lead shield was being assembled. The dust collection spread over a few months in the summer and fall of 2019 and 2021. This study provides insights regarding dust background mitigation procedures in place at SNOLAB, and informs backgrounds from particulate fallout during regular and specific laboratory activities.

\section{The SNOLAB facility and monitored locations}

SNOLAB \cite{smith2012snolab} is an internationally-recognized facility for deep underground science, located 2 km underground in the Vale Creighton mine, near Sudbury, Ontario. The science areas addressed by the experiments at SNOLAB primarily focus on subatomic and astroparticle physics questions, namely the search for galactic dark matter and the study of neutrino properties and sources. Both of these fields require an ultra-low radioactive background environment to attain target sensitivities. Radioactive backgrounds from cosmic radiation, material intrinsic radioactivity, and radioactivity from dust particulate fallout can hide or mimic searched signals, limiting detector sensitivities. The entire SNOLAB underground facility is a class 2000 cleanroom (corresponding to a cleanliness level between ISO 6 and ISO 7), providing an ideal low-background environment for rare-event studies. Since the facility is hosted in an active mine, the challenges of maintaining the cleanroom laboratory  are not trivial. To access the laboratory, both humans and materials have to pass through a vestibule area split into a \emph{dirty} side and a \emph{clean} side. Materials are unwrapped and cleaned at both sides before entering the facility. 
Everyone entering the underground facility has to take a full shower and don clean clothes and cleanroom garments. The facility also includes a surface building with office spaces, meeting rooms, and a class 2000 cleanroom laboratory including three research areas, a chemical laboratory and one assembly area. 

\begin{figure}[htb!]
    \centering
    \includegraphics[width=0.8\textwidth]{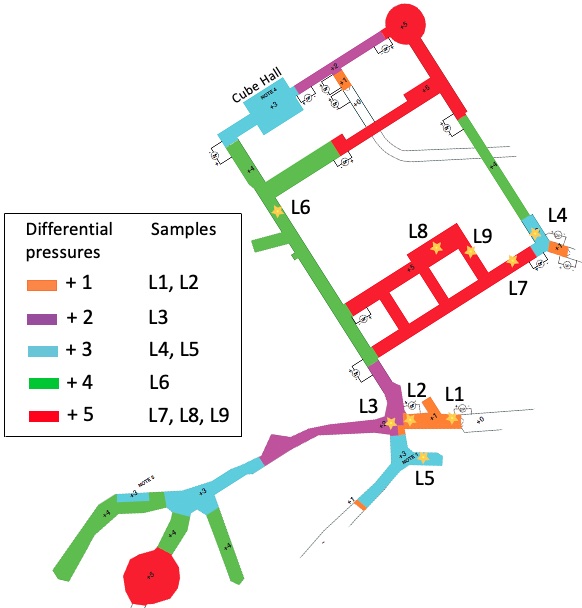}
    \caption{Map of the SNOLAB underground facility. Investigated locations are highlighted with a star. Locations are defined as L1: Rm 104, Dirty Carwash; L2: Rm 105, Clean Carwash; L3: Rm 123, Junction; L4: Rm 145, Machine Shop; L5: Rm 116, Refuge; L6: Drift J; L7: Rm 132, Chemistry Lab; L8: CUTE monorail; L9: SuperCDMS Utility Drift. A legend for the differential pressure color code is included. Locations monitored in each pressure area are also listed.}\label{fig:map}
\end{figure}

Nine selected underground locations and one surface location were monitored over a 60-day period between September and November 2019, and between July and September 2021, respectively. Figure~\ref{fig:map} shows the SNOLAB underground facility layout, along with the monitored locations tagged with a star and a number notation.

Part of the strategy for keeping the laboratory clean is to establish different pressure zones to force air from the cleanest spaces intended for the experiments to the dirtiest spaces (ultimately the exterior of the cleanroom laboratory).  The different pressure zones in the laboratory are shown in Fig.~\ref{fig:map}, where a color and a number is associated to each pressure zone. Higher number indicates higher pressure. 
The nominal design for the pressure differences between zones is 0.05” water column or 12.5 Pa. The legend in Figure \ref{fig:map} lists the pressure zones and the locations monitored in each zone.
An exception to keeping the laboratory spaces at the highest pressure is the Cube Hall (cyan colored square surface at the back of the underground laboratory in Figure~\ref{fig:map}) where there is presently one cryogenic liquid argon experiment. 
The Cube Hall is kept at a negative pressure relative to adjoining areas for safety. In case of a cryogenic leak into the main hall, the negative pressure will help contain the escaping inert gases within the hall.

The dirtiest space is the Carwash: the access point for materials to enter the underground facility. It is divided into a dirty and a clean area (L1 and L2 in Figure \ref{fig:map}, respectively) separated by doors. The Junction (L3) is the area linking the Carwash to the experimental areas and is highlighted in purple. Locations L4 and L5 are the Machine Shop and Refuge/Lunchroom areas, respectively. SNOLAB users and staff take their breaks, lunch, and snacks in the Refuge, where activities unusual for a cleanroom space are performed daily, including the use of coffee and tea machines, toasters, and microwave ovens.
Although locations L1, L2, L5 and L4 are thoroughly cleaned daily, they are not maintained at the same cleanroom particulate levels of the other locations.

\section{Experimental methodology}

\begin{figure}[htb!]
    \centering
    \includegraphics[width=0.5\textwidth]{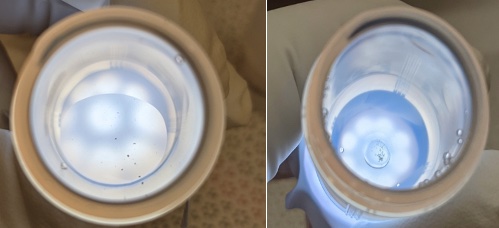}
    \caption{Samples with undissolved or partially dissolved particulate, collected in the SuperCDMS utility drift (L9, left), and the clean side of the car wash (L2, right).}
    \label{fig:particulate}
\end{figure}

Dust sample collection and dissolution was performed in ULB perfluoroalkoxy alkane (PFA) screw cap vials, following the method described in \cite{di2021direct}. PFA is a very clean, acid-resistant, low-background material, hence the reason for its use for this work. Vials were exposed over a time period of a month or longer. 
Particulate dissolution and analyses were performed at Pacific Northwest National Laboratory (PNNL). An Agilent 8900 triple quadrupole ICP-MS (Agilent Technologies, Santa Clara, CA), located in a class 10,000 cleanroom and equipped with an integrated autosampler, a microflow PFA nebulizer and a quartz double pass spray chamber, was used for the analyses. The analytical method developed and described in \cite{di2021direct} was followed, expanding the range of analytes to better assess how effective the mitigation strategies in place at SNOLAB are in keeping the mine dust out of the laboratory. Significant stable major constituents of the local rock \cite{SNOLAB} were included. In particular, we focused on Mg, Mn, Ni, and Zn. For each location, multiple replicates were collected. The number of replicates per location ranged from two to five, and are listed in Tables \ref{tab:bigmicroBq} and \ref{tab:pg} of the Appendix (section \ref{sec:appendix}). Some of the samples showed visible particulate deposited in some or all of the replicates, as reported in Tables \ref{tab:bigmicroBq} and \ref{tab:pg}. For these samples, the methodology was adapted accordingly, in order to fully dissolve the particulate prior to ICP-MS analysis. Concentrated solutions of Optima Grade HNO$_3$ ($\sim$50$\%$), HCl ($\sim$70$\%$), and HF ($\sim$20$\%$) were used to prepare acid mixtures for sample dissolution on a hotplate at 80$^o$C. The dissolution involved multiple cycles of sample treatment on the hotplate with HNO$_3$/HF and HNO$_3$/HCl mixtures. The percentage of HF and HCl in each mixture ranged from 10$\%$ to 40$\%$. Samples were treated on the hotplate for a few hours. Particulate fully dissolved in most of the samples, except for samples collected in the Carwash locations (L1 and L2), one out of five replicates collected in the Refuge room (L5), one replicate from the Drift area (L6), two replicates in L8, and all samples collected in L9. For samples with undissolved or partially dissolved particulate, undissolved particulate was excluded and solutions were analyzed on the ICP-MS. Figure \ref{fig:particulate} shows two samples from the SuperCDMS Utility Drift (L9) and clean side of the Carwash (L2) with undissolved or partially dissolved particulate. Details regarding the origin of the particulate are discussed in the following section.

\section{Results and discussions}\label{sec:results}

\begin{figure}[htb]
    \centering
    \includegraphics[width=0.95\textwidth]{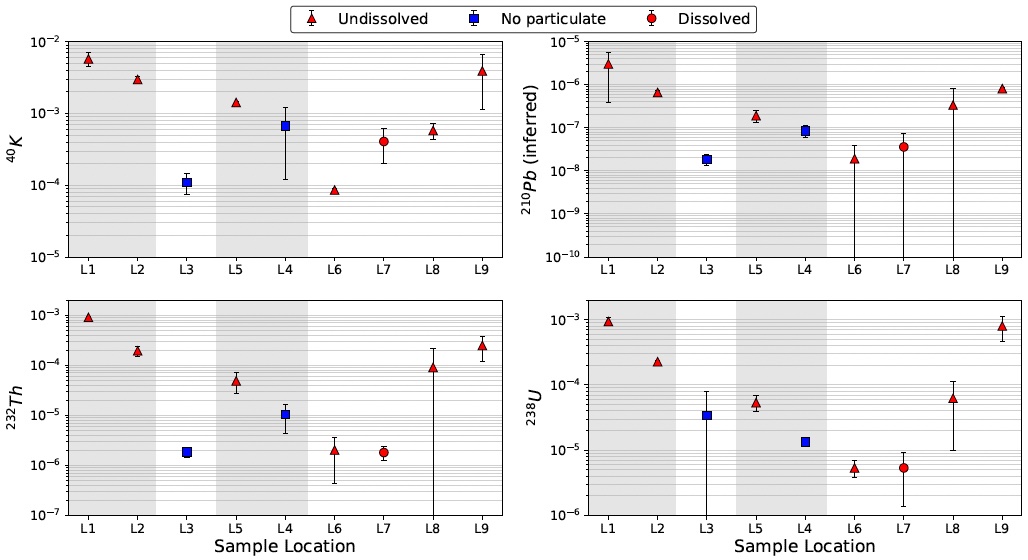}
    \caption{Accumulation rate of $^{40}$K, $^{210}$Pb, $^{232}$Th, and $^{238}$U, in $\mu$Bq$\cdot$day$^{-1}$$\cdot$cm$^{-2}$, measured in the underground locations. Data are labelled according to whether visible particulate was found in the collection vials, and, if present, whether it was fully dissolved or remained partially undissolved.}
    \label{fig:KPbThUunder}
\end{figure}

\begin{figure}[htb!]
    \centering
    \includegraphics[width=0.95\textwidth]{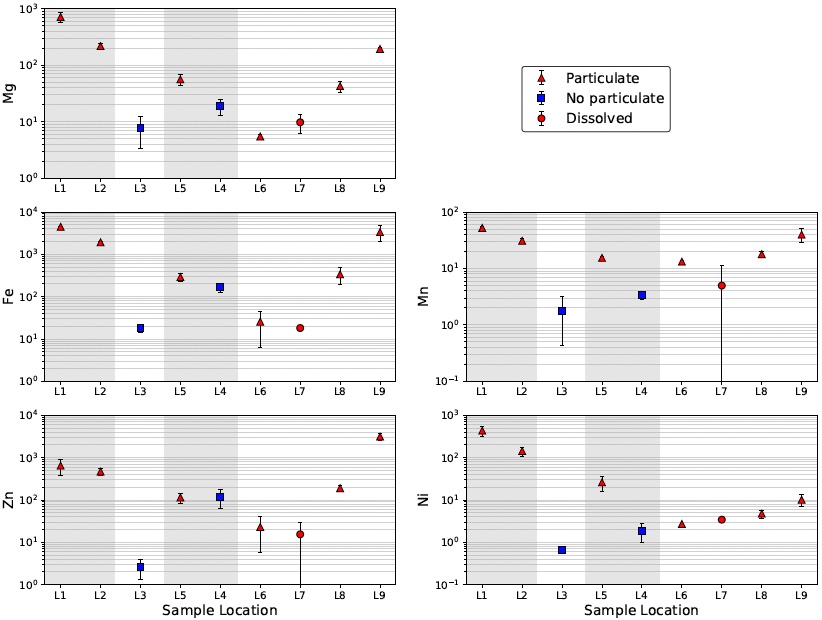}
    \caption{Accumulation rate of selected major elements in the local soil in pg$\cdot$day$^{-1}$$\cdot$cm$^{-2}$. Data are labelled according to whether visible particulate was found in the collection vials, and, if present, whether it was fully dissolved or remained partially undissolved.}
    \label{fig:others_under}
\end{figure}

Results for $^{40}$K, $^{210}$Pb, $^{232}$Th, and $^{238}$U fallout rates obtained in the underground locations are listed in Table \ref{tab:bigmicroBq} and plotted in Figure \ref{fig:KPbThUunder}. Accumulation rates measured for Mg, Mn, Fe, Ni, and Zn are listed in Table \ref{tab:pg} and shown in Figure \ref{fig:others_under}. In all plots and tables, locations are listed according to the relative pressure maintained in the room, from the lowest to the highest. 
Shaded areas in the plots indicate locations not maintained at cleanroom levels. Location L5 (the refuge station/lunchroom) is normally at the same pressure as the junction area (location L3), and it is not maintained at a cleanroom level. For this reason, L5 is listed in the plots right after L3, before L4. Tables \ref{tab:bigmicroBq} and \ref{tab:pg} list results for all replicates; for each replicate it is specified whether or not visible particulate was observed, and if observed, whether or not it was dissolved or remained partially undissolved. Plots in Figures \ref{fig:KPbThUunder} and \ref{fig:others_under} show results as the average and standard deviation of replicates in each location. Replicates for which accumulation rates were measured below limits of detection (LOD) were not included in the calculations of the plotted averages.

Results for $^{40}$K, $^{210}$Pb, $^{232}$Th, and $^{238}$U are reported in $\mu$Bq$\cdot$day$^{-1}\cdot$cm$^{-2}$. In order to determine accumulation rates in terms of activities, the same method previously adopted in \cite{di2021direct} was used. Activities were calculated from the masses measured via ICP-MS based on the specific activity of each radionuclide and natural isotopic abundance. For $^{210}$Pb, a reference value of 200 Bq of $^{210}$Pb per kg of stable Pb from a previous measurement of Pb bricks \cite{alessandrello1998measurements} was assumed. The same value had been used in \cite{di2021direct} to infer radioactivity of $^{210}$Pb from measured stable Pb. Isotopes $^{206}$Pb, $^{207}$Pb, and $^{208}$Pb were monitored for the ICP-MS determination of stable Pb. Accumulation rates measured for Mg, Mn, Fe, Ni, and Zn are reported, in pg$\cdot$day$^{-1}$$\cdot$cm$^{-2}$.

It is evident that the measures in place at SNOLAB to mitigate mine dust entrance in the laboratory are extremely effective, especially looking at the accumulation rates of isotopes such as Ni, Mn, Zn and Mg which are reduced by two to three orders of magnitude from the dirty side of the Carwash (L1) to the Junction area (L3). 
The mitigation trend in accumulation rates is consistent for all the measured isotopes. Variations from the trend are observed in the cleanest experimental areas, most likely due to visible particulate collected in some of the replicates. In particular, locations L8 and L9 which refer to the Cryogenic Underground TEst facilty (CUTE) deck at 3 m height and the SuperCDMS Utility Drift, reported visible particulate (shown in Figure \ref{fig:particulate}) in all the vials exposed in these locations. 
For the latter location, the door between the main experimental areas and the Utility Drift was being assembled. The door is made of a bulk of fire-retardant and acoustic foam in a stainless steel case. The foam was cut to fit the case underground in the SuperCDMS Utility Drift area. The particulate found in the vials (left picture in Figure \ref{fig:particulate}) looked similar to the foam.
At the CUTE deck, the vials were located on top of the post of the monorail gantry crane which is used to move the CUTE cryostat in and out the water tank and shielding, and the cleanroom. The particulate found in the vials at CUTE was consistent with shredded coating of the monorail beam during the crane usage. The presence of particulates explains the larger error bars accumulation rates for these locations in Fig.~\ref{fig:KPbThUunder} and the increasing fallout rates in the cleanest experimental areas.

\begin{figure}[htb]
    \centering
    \includegraphics[width=0.55\textwidth]{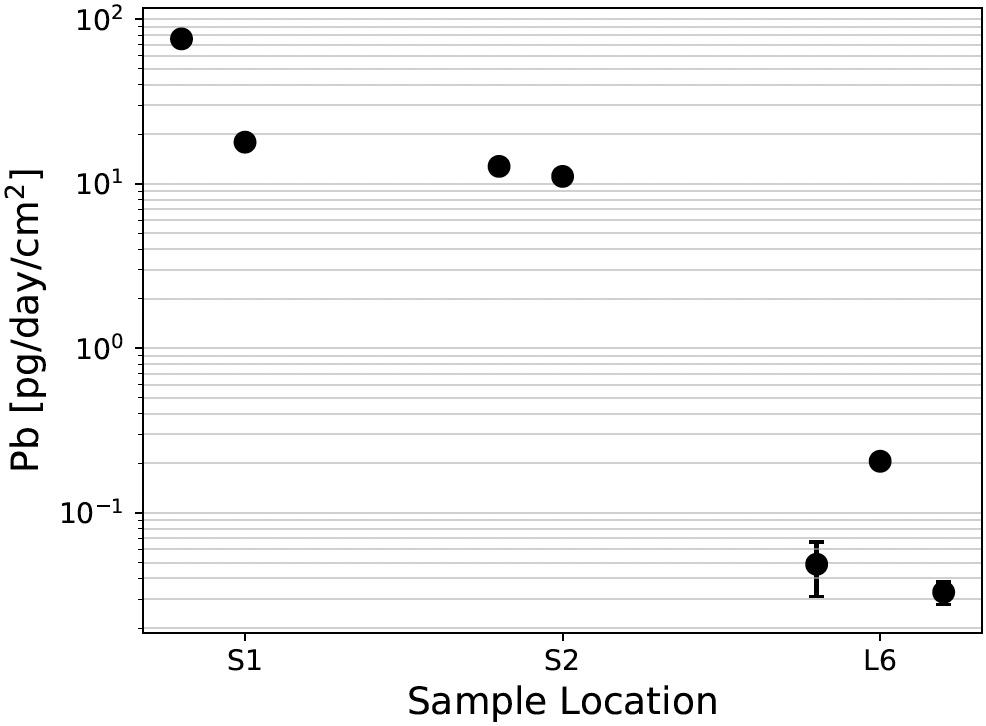}
    \caption{Accumulation rate of stable Pb, in pg$\cdot$day$^{-1}$$\cdot$cm$^{-2}$, measured in L6 during regular laboratory activities and in two nearby locations (S1 and S2) during handling of Pb bricks. Error bars are hidden by the markers where not visible.}
    \label{fig:Pb_plot}
\end{figure}

Location L6, the so called J-Drift hosting charge coupled device (CCD) experiments, was also monitored over an additional 6-month period in 2021, during the assembly of a lead shield and water tanks for an experiment. Two locations were chosen around the area of the shielding assembly, which we will refer to as S1 and S2. Two replicates for each of these locations were acquired. 
Accumulation rates for stable Pb, in pg$\cdot$day$^{-1}\cdot$cm$^{-2}$, measured in all replicates of location L6 (exposure during no lead activity ) and S1 and S2 (exposure during lead shielding assembly) are plotted in Fig \ref{fig:Pb_plot} and listed in Table~\ref{tab:sensei_Pb}. 

Fallout rates of stable Pb during Pb bricks handling (S1 and S2) were measured two to three orders of magnitude higher compared to fallout rates obtained while no activity involving Pb bricks was being performed (L6).
Fallout rates of isotopes $^{40}$K, $^{232}$Th and $^{238}$U were also monitored in locations S1 and S2. Results show no significant variation in accumulation rates for those isotopes while handling lead bricks. Data are reported in Table \ref{tab:L6S1S2_other} as the average and standard deviation of replicates, along with data taken in that location when no lead was being handled (L6).

The Pb shield that was assembled underground is composed of double chevron bricks for a total mass of about 4.3~tonnes. A sample of the Pb used for the shielding was screened underground at SNOLAB via a high purity germanium (HPGe) gamma spectroscopy detector \cite{lawson2020low}. The counted sample was 56.5~g and placed in the standard counting vial in the well of the HPGe detector. The detector is a 300 cm$^{3}$ p-type Canberra Well HPGe within a cylindrical shielding of 2 inches of Cu and 8 inches of Pb. The $^{210}$Pb measurement gave a result of 4.2 $\pm$ 1.0~ Bq$\cdot$kg$^{-1}$ analyzing the 46~keV line.
\color{black}Based on this measurement, the fallout rate of $^{210}$Pb from the lead bricks in S1 and S2 can be calculated as (2.0 $\pm$ 1.8)$\times$10$^{-10}$ and (5.0 $\pm$ 1.3)$\times$10$^{-11}$ $\mu$Bq$\cdot$day$^{-1}$$\cdot$cm$^{-2}$, respectively. It is worth pointing out that these estimated fallout rates in activity are assuming that the fallout rate of stable Pb from the bricks is the only source of $^{210}$Pb, and does not account for any other source of $^{210}$Pb. These estimates can be easily scaled to the mass of the lead shield and lead purity considered by each experiment to better assess the radioactive contamination contribution in terms of $^{210}$Pb dust from Pb shielding, which could contribute to the radioactive background budget of the experiment.  
The location was also monitored for Mg, Mn, Fe, Ni, and Zn during the lead shield construction. Results are listed in Table \ref{tab:S_others} of Section \ref{sec:appendix}. Results are compatible with those previously obtained in the same location (L6 in Table \ref{tab:pg}), with the exception of Fe, which showed higher accumulation rates in those replicates of L6 where visible particulate was observed. The higher accumulation rates of Fe can be attributed to the presence of the particulate.

\begin{table}
\centering
\begin{tabular}{| c | c  c c |} 
\hline 
 & \multicolumn{3}{|c|}{Accumulation rate [$\mu$Bq$\cdot$day$^{-1}\cdot$cm$^{-2}$]} \\
 \cline{2-4}
  & $^{40}$K & $^{232}$Th & $^{238}$U \\
\hline
L6 & (8.5 $\pm$ 0.4)$\cdot$ 10$^{-5}$ & (2.0 $\pm$ 1.6)$\cdot$ 10$^{-6}$ & (5.3 $\pm$ 1.5)$\cdot$ 10$^{-6}$\\  
S1 & (1.0 $\pm$ 0.3)$\cdot$ 10$^{-4}$ & (1.7 $\pm$ 0.8)$\cdot$ 10$^{-7}$& (1.6 $\pm$ 0.7)$\cdot$ 10$^{-6}$\\
S2 & (7 $\pm$ 2)$\cdot$ 10$^{-5}$ & (1.8 $\pm$ 0.1)$\cdot$ 10$^{-7}$& (5.2 $\pm$ 0.2)$\cdot$ 10$^{-7}$ \\
\hline
\end{tabular}
\caption{Fallout rate of $^{40}$K $^{232}$Th $^{238}$U, in $\mu$Bq$\cdot$day$^{-1}$$\cdot$cm$^{-2}$, measured in the two S1 and S2 locations during the shield construction. For comparison, fallout rates measured in the nearby L6 location, measured in a regular activity time, are also listed.}
\label{tab:L6S1S2_other}
\end{table}

In 2021, two locations at the SNOLAB surface cleanroom laboratory were also monitored. A set of vials was placed at the hallway giving access to a single cleanroom and chemistry laboratory. A second set of vials was placed at the clean assembly area, which is the cleanest space at the surface cleanroom facility. The clean assembly area is the area where leaching, etching, and passivation of shielding and detector materials take place, as well as detector assembly. It is the only room of the SNOLAB surface facility equipped with a fixed dust laser detector to monitor particulates rates. 
Measurements in terms of fallout rate of $^{40}$K, $^{210}$Pb, $^{232}$Th $^{238}$U, in $\mu$Bq$\cdot$day$^{-1}$$\cdot$cm$^{-2}$ are listed in Tab.~\ref{tab:surface}. The accumulation rates have been compared to the underground cleanest location (L6). %of the dust collection campaign of 2019, location L6, the J drift. 
There is no significant difference between fallout rates from the surface clean spaces and the underground cleanroom location.

\color{black}

\begin{table}
\resizebox{\textwidth}{!}{
%\centering
\begin{tabular}{|c c| c c c c |} 
\hline 
 &  &\multicolumn{4}{|c|}{Accumulation rate [$\mu$Bq$\cdot$day$^{-1}\cdot$cm$^{-2}$]}\\
 \cline{3-6}
 & & $^{40}$K & $^{210}$Pb & $^{232}$Th & $^{238}$U\\
\hline
\multirow{2}{*}{\textbf{Surface}} & Hallway & (9.3 $\pm$ 1.0)$\cdot$ 10$^{-5}$ & (1.9 $\pm$ 1.9)$\cdot$ 10$^{-8}$ & (1.4 $\pm$ 0.2)$\cdot$ 10$^{-6}$ & (2.7 $\pm$ 0.6)$\cdot$ 10$^{-6}$\\  
& Clean assembly & (4 $\pm$ 3)$\cdot$ 10$^{-5}$ & (6 $\pm$ 4)$\cdot$ 10$^{-10}$& (2.4 $\pm$ 2.4)$\cdot$ 10$^{-7}$ & (1.4 $\pm$ 2.3)$\cdot$ 10$^{-6}$\\
\hline
\textbf{Underground} & L6 & (8.5 $\pm$ 0.4)$\cdot$ 10$^{-5}$ & (1.9 $\pm$ 1.9)$\cdot$ 10$^{-8}$& (2.0 $\pm$ 1.6)$\cdot$ 10$^{-6}$ & (5.2 $\pm$ 0.2)$\cdot$ 10$^{-7}$ \\
\hline
\end{tabular}}
\caption{Fallout rate of $^{40}$K, $^{210}$Pb, $^{232}$Th $^{238}$U, in $\mu$Bq$\cdot$day$^{-1}$$\cdot$cm$^{-2}$, measured in two locations of the surface cleanroom. For comparison, fallout rates measured in the L6 underground location are also listed.}
\label{tab:surface}
\end{table}

\section{Conclusions}

\color{black}

This study demonstrates our method, described in \cite{di2021direct}, as an effective tool to inform radioactive backgrounds from dust particulate on ULB materials during routine and/or specific laboratory activites, and evaluate the efficacy of dust mitigation procedures. Key locations were monitored at SNOLAB, a one-of-a-kind, internationally-recognized, underground facility hosting rare-event physics experiments, where dust control is crucial. Results obtained from the monitoring of major stable constituents of the local rock and Ni (from the mine hosting the laboratory) allowed us to evaluate and demonstrate the efficacy of mitigation procedures in place to keep the mine dust out of experimental areas. A two to three order of magnitude decrease in fallout rates of stable elements from the rock and of Ni from the mine was observed between the Carwash area, the transition area from the mine to the laboratory, and the clean Junction area inside the laboratory. The monitoring of radionuclides of concern for ULB detectors showed a correlation between specific laboratory activities and increased fallout rates, and demonstrated that specific laboratory activities can increase fallout rates by orders of magnitudes. In particular, an increase of one to two orders of magnitude in the fallout rates of $^{40}$K, $^{210}$Pb, $^{232}$Th, and $^{238}$U was observed in clean experimental areas (L8 and L9) where specific laboratory and assembly operations were being performed, compared to other clean experimental areas maintained at the same level of cleanliness (e.g., L6 and L7). The presence of particulate compatible with materials handled in those location during sample collection validated the correlation between specific activities being performed and increased fallout rates of radionuclides. A full dissolution of the undissolved particulate may reveal even further increased fallout rates. A two to three order of magnitude increase in stable Pb fallout rates was observed in the the underground location monitored during operations involving handling of Pb bricks. The compatibility between data obtained in the underground clean spaces and the clean areas in the surface building further demonstrated the efficacy of the procedures in place at SNOLAB in keeping the mine dust out of the underground laboratory.

This study validates the dust monitoring methodology as a valuable tool to inform backgrounds and evaluate mitigation techniques. In addition, results confirm that dust particulate composition in cleanrooms is strongly determined by ongoing activities and handled materials. Assumed dust composition in cleanrooms can therefore lead to inaccurate background predictions, as already found in our previous study \cite {di2021direct}. 

Going forward, methods will be developed to directly determine radioactive background contributions from $^{238}$U progeny (e.g., $^{210}$Pb and $^{226}$Ra) in dust for further improved background predictions.

\section{Acknowledgements}

Pacific Northwest National Laboratory (PNNL) is operated by Battelle for the United States Department of Energy (DOE) under Contract no. DE-AC05-76RL01830. This study was supported by the DOE Office of High Energy Physics Advanced Technology R\&D subprogram.
The authors would like to thank SNOLAB and its staff for support through underground space, logistical and technical services. SNOLAB operations are supported by the Canada Foundation for Innovation and the Province of Ontario Ministry of Research and Innovation, with underground access provided by Vale at the Creighton mine site.

\bibliographystyle{elsarticle-num} 
\bibliography{references.bib}

%\newpage
\section{Appendix}\label{sec:appendix}

\begin{table}
\resizebox{\textwidth}{!}{
%\centering
\begin{tabular}{| c | c | c | c c c c |} 
\hline 
 Location & Replicate & Particulate & \multicolumn{4}{|c|}{Accumulation rate [$\mu$Bq$\cdot$day$^{-1}\cdot$cm$^{-2}$]}\\
 \hline
& & & $^{40}$K & $^{210}$Pb & $^{232}$Th & $^{238}$U\\
\hline
L1 & 1 & Undissolved & (6.6 $\pm$ 0.1)$\times$10$^{-5}$ & (4.97 $\pm$ 0.04)$\times$10$^{-8}$ & (8.9 $\pm$ 0.3)$\times$10$^{-4}$ & (1.03 $\pm$ 0.07)$\times$10$^{-3}$\\  
 & 2 & Undissolved & (4.9 $\pm$ 0.1)$\times$10$^{-5}$ & (1.17 $\pm$ 0.01)$\times$10$^{-6}$ & (9.2 $\pm$ 0.2)$\times$10$^{-4}$ & (8.5 $\pm$ 0.9)$\times$10$^{-4}$\\  
\hline
L2 & 1 & Undissolved & (2.8 $\pm$ 0.1)$\times$10$^{-3}$ & (7.2 $\pm$ 0.1)$\times$10$^{-7}$ & (2.0 $\pm$ 0.1)$\times$10$^{-4}$ & (2.3 $\pm$ 0.3)$\times$10$^{-4}$\\ 
 & 2 & Undissolved & (2.8 $\pm$ 0.1)$\times$10$^{-3}$ & (5.8 $\pm$ 0.1)$\times$10$^{-7}$ & (1.5 $\pm$ 0.1)$\times$10$^{-4}$ & (2.2 $\pm$ 0.2)$\times$10$^{-4}$\\ 
 & 3 & Undissolved & (3.4 $\pm$ 0.1)$\times$10$^{-3}$ & (6.85 $\pm$ 0.03)$\times$10$^{-7}$ & (2.3 $\pm$ 0.1)$\times$10$^{-4}$ & (2.3 $\pm$ 0.2)$\times$10$^{-4}$\\ 
 \hline
 L3 & 1 & Not observed & (6.9 $\pm$ 0.4)$\times$10$^{-5}$ & (2.5 $\pm$ 0.3)$\times$10$^{-8}$ & $<$ 4.9$\times$10$^{-7}$ &  $<$ 1.1$\times$10$^{-6}$\\  
 & 2 & Not observed & (1.27 $\pm$ 0.04)$\times$10$^{-4}$ & (1.6 $\pm$ 0.3)$\times$10$^{-8}$ & (1.6 $\pm$ 0.5)$\times$10$^{-6}$ & (6.5 $\pm$ 1.1)$\times$10$^{-5}$\\ 
 & 3 & Not observed & (1.34 $\pm$ 0.04)$\times$10$^{-4}$ & (1.5 $\pm$ 0.2)$\times$10$^{-8}$ & (2 $\pm$ 2)$\times$10$^{-6}$ & (2.9 $\pm$ 1.5)$\times$10$^{-6}$\\ 
 \hline
 L5 & 1 & Dissolved & (1.4 $\pm$ 0.1)$\times$10$^{-3}$ & (2.00 $\pm$ 0.01)$\times$10$^{-7}$ & (4.0 $\pm$ 0.9)$\times$10$^{-5}$ & (4.3 $\pm$ 1.6)$\times$10$^{-5}$\\ 
 & 2 & Undissolved & (1.3 $\pm$ 0.1)$\times$10$^{-3}$ & (2.64 $\pm$ 0.05)$\times$10$^{-7}$ & (3.1 $\pm$ 0.2)$\times$10$^{-5}$ & (6.4 $\pm$ 1.4)$\times$10$^{-5}$\\ 
 & 3 & Dissolved & (1.5 $\pm$ 0.1)$\times$10$^{-3}$ & (1.80 $\pm$ 0.03)$\times$10$^{-7}$ & (8.7 $\pm$ 1.5)$\times$10$^{-5}$ & (7.3 $\pm$ 0.7)$\times$10$^{-5}$\\ 
 & 4 & Dissolved & (1.5 $\pm$ 0.1)$\times$10$^{-3}$ & (2.2 $\pm$ 0.1)$\times$10$^{-7}$ & (4.1 $\pm$ 0.8)$\times$10$^{-5}$ & (4.7 $\pm$ 1.1)$\times$10$^{-5}$\\ 
 & 5 & Dissolved & (1.4 $\pm$ 0.1)$\times$10$^{-3}$ & (9.8 $\pm$ 0.1)$\times$10$^{-8}$ & (4.4 $\pm$ 0.4)$\times$10$^{-5}$ & (4 $\pm$ 2)$\times$10$^{-5}$\\  
 \hline
L4 & 1 & Not observed & (4.1 $\pm$ 0.1)$\times$10$^{-4}$ & (1.20 $\pm$ 0.04)$\times$10$^{-7}$ & (1.2 $\pm$ 0.2)$\times$10$^{-5}$ & (1.4 $\pm$ 1.0)$\times$10$^{-5}$\\  
 & 2 & Not observed & (3.55 $\pm$ 0.05)$\times$10$^{-4}$ & (9.1 $\pm$ 0.4)$\times$10$^{-8}$ & (7 $\pm$ 2)$\times$10$^{-6}$ & (1.1 $\pm$ 0.4)$\times$10$^{-5}$\\  
 & 3 & Not observed & (4.13 $\pm$ 0.04)$\times$10$^{-4}$ & (6.7 $\pm$ 0.3)$\times$10$^{-8}$ & (1.8 $\pm$ 0.7)$\times$10$^{-5}$ & (1.4 $\pm$ 1.1)$\times$10$^{-5}$\\  
 & 4 & Not observed & (1.48 $\pm$ 0.01)$\times$10$^{-3}$ & (6.5 $\pm$ 0.2)$\times$10$^{-8}$ & (4.1 $\pm$ 0.5)$\times$10$^{-6}$ & (1.5 $\pm$ 0.5)$\times$10$^{-5}$\\  
 \hline
L6  & 1 & Not observed & (8.5 $\pm$ 0.4)$\times$10$^{-5}$ & (10 $\pm$ 4)$\times$10$^{-9}$ & (4 $\pm$ 5)$\times$10$^{-6}$ & (5.4 $\pm$ 1.5)$\times$10$^{-6}$\\ 
  & 2 & Undissolved & $<$ 2.0$\times$10$^{-4}$ & (4.1 $\pm$ 0.3)$\times$10$^{-8}$ & (1.4 $\pm$ 1.3)$\times$10$^{-6}$ &  $<$ 6.2$\times$10$^{-6}$\\ 
 & 3 & Dissolved & $<$ 2.0$\times$10$^{-4}$ & (6.6 $\pm$ 1.0)$\times$10$^{-9}$ & (8 $\pm$ 5)$\times$10$^{-7}$ &  $<$ 6.2$\times$10$^{-6}$\\ 
 \hline
L7  & 1 & Not observed & (4.1 $\pm$ 0.1)$\times$10$^{-4}$ & (1.5 $\pm$ 0.3)$\times$10$^{-8}$ & $<$ 5.0$\times$10$^{-7}$ & (5 $\pm$ 8)$\times$10$^{-6}$\\  
 & 2 & Dissolved & (7.0 $\pm$ 0.7)$\times$10$^{-4}$ & (9.3 $\pm$ 0.2)$\times$10$^{-8}$ & (1.7 $\pm$ 1.0)$\times$10$^{-6}$ & (1.1 $\pm$ 0.6)$\times$10$^{-5}$\\ 
  & 3 & Not observed & (2.3 $\pm$ 0.1)$\times$10$^{-4}$ & (2.9 $\pm$ 0.3)$\times$10$^{-8}$ & (1.3 $\pm$ 1.8)$\times$10$^{-6}$ & (4 $\pm$ 3)$\times$10$^{-6}$\\ 
 & 4 & Not observed & (2.95 $\pm$ 0.04)$\times$10$^{-4}$ & (7.8 $\pm$ 1.7)$\times$10$^{-9}$ & (2.4 $\pm$ 1.3)$\times$10$^{-6}$ & (2 $\pm$ 4)$\times$10$^{-6}$\\   
 \hline
 L8  & 1 & Dissolved & (5.5 $\pm$ 0.7)$\times$10$^{-4}$ & (7.5 $\pm$ 0.3)$\times$10$^{-8}$ & (4.1 $\pm$ 0.1)$\times$10$^{-5}$ & (4.3 $\pm$ 0.7)$\times$10$^{-5}$\\  
 & 2 & Undissolved & (8.0 $\pm$ 0.8)$\times$10$^{-4}$ & (1.54 $\pm$ 0.01)$\times$10$^{-7}$ & (3.3 $\pm$ 0.1)$\times$10$^{-5}$ & (5.3 $\pm$ 1.6)$\times$10$^{-5}$\\  
  & 3 & Undissolved & (4.9 $\pm$ 0.7)$\times$10$^{-4}$ & (1.04 $\pm$ 0.02)$\times$10$^{-6}$ & (2.76 $\pm$ 0.04)$\times$10$^{-4}$ & (1.4 $\pm$ 0.1)$\times$10$^{-5}$\\  
 & 4 & Dissolved & (4.8 $\pm$ 0.7)$\times$10$^{-4}$ & (8.88 $\pm$ 0.06)$\times$10$^{-8}$ & (1.2 $\pm$ 0.5)$\times$10$^{-5}$ & (1.6 $\pm$ 0.4)$\times$10$^{-5}$\\  
 \hline
L9  & 1 & Undissolved & (5.9 $\pm$ 0.1)$\times$10$^{-3}$ & (7.68 $\pm$ 0.03)$\times$10$^{-7}$ & (3.4 $\pm$ 0.2)$\times$10$^{-4}$ & (1.04 $\pm$ 0.04)$\times$10$^{-3}$\\   
 & 2 & Undissolved & (2.0 $\pm$ 0.1)$\times$10$^{-3}$ & (8.67 $\pm$ 0.03)$\times$10$^{-7}$ & (1.6 $\pm$ 0.2)$\times$10$^{-4}$ & (5.7 $\pm$ 0.4)$\times$10$^{-4}$\\ 
\hline
\end{tabular}}
\caption{Measured fallout rate of $^{40}$K, $^{210}$Pb, $^{232}$Th $^{238}$U, in $\mu$Bq$\cdot$day$^{-1}$$\cdot$cm$^{-2}$, in each replicate of the monitored underground locations. For each replicate, information is provided about whether or not visible particulate was observed, and if observed, whether or not it was dissolved or part of it remained undissolved.}
\label{tab:bigmicroBq}
\end{table}

\begin{table}
\resizebox{\textwidth}{!}{
%\centering
\begin{tabular}{| c | c | c | c c c c c |} 
\hline 
 Location & Replicate & Particulate & \multicolumn{5}{|c|}{Accumulation rate [pg$\cdot$day$^{-1}\cdot$cm$^{-2}$]}\\
 \hline
& & & Mg & Mn & Fe & Ni & Zn\\
\hline
L1 & 1 & Undissolved & (8.3 $\pm$ 0.3)$\times$10$^{2}$ & (5.5 $\pm$ 0.2)$\times$10$^{1}$ & (4.60 $\pm$ 0.01)$\times$10$^{3}$ & (5.2 $\pm$ 0.1)$\times$10$^{2}$ & (8.4 $\pm$ 0.2)$\times$10$^{2}$\\  
 & 2 & Undissolved & (6.1 $\pm$ 0.1)$\times$10$^{2}$ & (5.0 $\pm$ 0.1)$\times$10$^{1}$ & (4.40 $\pm$ 0.04)$\times$10$^{3}$ & (3.5 $\pm$ 0.1)$\times$10$^{2}$ & (4.6 $\pm$ 0.1)$\times$10$^{2}$\\  
\hline
L2 & 1 & Undissolved & (2.4 $\pm$ 0.1)$\times$10$^{2}$ & (2.9 $\pm$ 0.2)$\times$10$^{1}$ & (2.00 $\pm$ 0.01)$\times$10$^{3}$ & (1.80 $\pm$ 0.1)$\times$10$^{2}$ & (4.0 $\pm$ 0.1)$\times$10$^{2}$\\ 
 & 2 & Undissolved & (1.9 $\pm$ 0.1)$\times$10$^{2}$ & (2.9 $\pm$ 0.2)$\times$10$^{1}$ & (1.80 $\pm$ 0.02)$\times$10$^{3}$ & (1.2 $\pm$ 0.3)$\times$10$^{2}$ & (4.4 $\pm$ 0.1)$\times$10$^{2}$\\ 
 & 3 & Undissolved & (2.3 $\pm$ 0.1)$\times$10$^{2}$ & (3.4 $\pm$ 0.2)$\times$10$^{1}$ & (1.90 $\pm$ 0.02)$\times$10$^{3}$ & (1.2 $\pm$ 0.2)$\times$10$^{2}$ & (5.7 $\pm$ 0.1)$\times$10$^{2}$\\ 
 \hline
 L3 & 1 & Not observed & (5.5 $\pm$ 1.4) & (1.8 $\pm$ 1.4) & (2.00 $\pm$ 0.04)$\times$10$^{1}$ & $<$ 6.7$\times$10$^{-1}$ & (4.0 $\pm$ 0.6) \\ 
 & 2 & Not observed & (1.3 $\pm$ 0.2)$\times$10$^{1}$ & $<$ 8.3$\times$10$^{-2}$ & (2.10 $\pm$ 0.04)$\times$10$^{1}$ & $<$ 6.7$\times$10$^{-1}$ & (2.2 $\pm$ 0.5) \\ 
 & 3 & Not observed & (4.9 $\pm$ 1.4) & $<$ 8.3$\times$10$^{-2}$ & (1.40 $\pm$ 0.04)$\times$10$^{1}$ & $<$ 6.7$\times$10$^{-1}$ & (1.5 $\pm$ 0.4) \\ 
 \hline
 L5 & 1 & Dissolved & (7.4 $\pm$ 0.3)$\times$10$^{1}$ & (1.6 $\pm$ 0.1)$\times$10$^{1}$ & (4.00 $\pm$ 0.02)$\times$10$^{2}$ & (2.00 $\pm$ 0.02)$\times$10$^{1}$ & (9.2 $\pm$ 0.2)$\times$10$^{1}$ \\ 
 & 2 & Undissolved & (6.2 $\pm$ 0.2)$\times$10$^{1}$ & (1.5 $\pm$ 0.1)$\times$10$^{1}$ & (3.00 $\pm$ 0.01)$\times$10$^{2}$ & (4.30 $\pm$ 0.03)$\times$10$^{1}$ & (1.6 $\pm$ 0.1)$\times$10$^{2}$ \\ 
 & 3 & Dissolved & (4.3 $\pm$ 0.1)$\times$10$^{1}$ & (1.5 $\pm$ 0.1)$\times$10$^{1}$ & (2.70 $\pm$ 0.01)$\times$10$^{2}$ & (1.7 $\pm$ 0.1)$\times$10$^{1}$ & (8.6 $\pm$ 0.1)$\times$10$^{1}$  \\ 
 & 4 & Dissolved & (5.2 $\pm$ 0.1)$\times$10$^{1}$ & (1.7 $\pm$ 0.1)$\times$10$^{1}$ & (2.60 $\pm$ 0.03)$\times$10$^{2}$ & (2.70 $\pm$ 0.04)$\times$10$^{1}$ & (1.20 $\pm$ 0.03)$\times$10$^{2}$ \\  
 & 5 & Dissolved & (5.0 $\pm$ 0.1)$\times$10$^{1}$ & (1.5 $\pm$ 0.1)$\times$10$^{1}$ & (2.30 $\pm$ 0.02)$\times$10$^{2}$ & (2.40 $\pm$ 0.02)$\times$10$^{1}$ & (1.20 $\pm$ 0.01)$\times$10$^{2}$ \\
 \hline
L4 & 1 & Not observed & (1.8 $\pm$ 0.1)$\times$10$^{1}$ & (4.0 $\pm$ 1.5) & (2.10 $\pm$ 0.02)$\times$10$^{2}$ & (3.0 $\pm$ 0.3) & (1.00 $\pm$ 0.01)$\times$10$^{2}$\\  
 & 2 & Not observed & (2.7 $\pm$ 0.2)$\times$10$^{1}$ & (2.7 $\pm$ 1.5) & (1.50 $\pm$ 0.01)$\times$10$^{2}$ & (1.2 $\pm$ 0.3) & (5.7 $\pm$ 0.1)$\times$10$^{1}$ \\  
 & 3 & Not observed & (1.4 $\pm$ 0.2)$\times$10$^{1}$ & (3.2 $\pm$ 1.4) & (1.30 $\pm$ 0.01)$\times$10$^{2}$ & (1.2 $\pm$ 0.3) & (1.30 $\pm$ 0.02)$\times$10$^{2}$\\  
 & 4 & Not observed & (1.7 $\pm$ 0.2)$\times$10$^{1}$ & (3.4 $\pm$ 1.4) & (1.80 $\pm$ 0.04)$\times$10$^{2}$ & (2.1 $\pm$ 0.4) & (1.90 $\pm$ 0.04)$\times$10$^{2}$\\  
 \hline
L6  & 1 & Not observed & $<$ 4.1 & $<$8.3$\times$10$^{-2}$ & (9.3 $\pm$ 0.4) & $<$ 6.7$\times$10$^{-1}$ & (3.9 $\pm$ 0.5) \\ 
  & 2 & Undissolved & (5.1 $\pm$ 0.7) & (1.4 $\pm$ 0.1)$\times$10$^{1}$ & (4.70 $\pm$ 0.05)$\times$10$^{1}$ &  (2.8 $\pm$ 0.3) & (3.7 $\pm$ 0.1)$\times$10$^{1}$ \\ 
 & 3 & Dissolved & (5.9 $\pm$ 0.9) & (1.3 $\pm$ 0.1)$\times$10$^{1}$ & (2.10 $\pm$ 0.04)$\times$10$^{1}$ &  (2.6 $\pm$ 0.2) & (2.8 $\pm$ 0.1)$\times$10$^{1}$\\ 
 \hline
L7  & 1 & Not observed & (6.0 $\pm$ 1.5) & (2.0 $\pm$ 1.8) & (1.90 $\pm$ 0.04)$\times$10$^{1}$ & $<$ 6.7$\times$10$^{-1}$ & (1.10 $\pm$ 0.04)$\times$10$^{1}$\\  
 & 2 & Dissolved & (1.0 $\pm$ 0.1) & (1.5 $\pm$ 0.2)$\times$10$^{1}$ & (1.70 $\pm$ 0.03)$\times$10$^{1}$ & (3.4 $\pm$ 0.4) & (3.7 $\pm$ 0.1)$\times$10$^{1}$ \\ 
  & 3 & Not observed & (1.3 $\pm$ 0.1)$\times$10$^{1}$ & (1.8 $\pm$ 1.7) & (1.80 $\pm$ 0.04)$\times$10$^{1}$ & $<$ 6.7$\times$10$^{-1}$ & (9.0 $\pm$ 1.2)\\ 
 & 4 & Not observed & $<$ 4.1 & (1.6 $\pm$ 1.5) & (1.90 $\pm$ 0.04)$\times$10$^{1}$ & $<$ 6.7$\times$10$^{-1}$ & (4.8 $\pm$ 0.4) \\   
 \hline
 L8  & 1 & Dissolved & (3.8 $\pm$ 0.2)$\times$10$^{1}$ & (1.7 $\pm$ 0.2)$\times$10$^{1}$ & (3.00 $\pm$ 0.06)$\times$10$^{2}$ & (5.9 $\pm$ 0.3) & (2.10 $\pm$ 0.03)$\times$10$^{2}$ \\  
 & 2 & Undissolved & (5.7 $\pm$ 0.2)$\times$10$^{1}$ & (2.1 $\pm$ 0.2)$\times$10$^{1}$ & (5.20 $\pm$ 0.01)$\times$10$^{2}$ & (5.0 $\pm$ 0.3) & (2.20 $\pm$ 0.05)$\times$10$^{2}$ \\  
  & 3 & Undissolved & (4.2 $\pm$ 0.2)$\times$10$^{1}$ & (1.7 $\pm$ 0.2)$\times$10$^{1}$ & (3.60 $\pm$ 0.01)$\times$10$^{2}$ & (4.7 $\pm$ 0.3) & (1.60 $\pm$ 0.02)$\times$10$^{2}$\\  
 & 4 & Dissolved & (3.4 $\pm$ 0.1)$\times$10$^{1}$ & (1.6 $\pm$ 0.2)$\times$10$^{1}$ & (1.80 $\pm$ 0.04)$\times$10$^{2}$ & (3.5 $\pm$ 0.2) & (1.70 $\pm$ 0.02)$\times$10$^{2}$ \\  
 \hline
L9  & 1 & Undissolved & (1.9 $\pm$ 0.1)$\times$10$^{2}$ & (4.7 $\pm$ 0.2)$\times$10$^{1}$ & (4.30 $\pm$ 0.01)$\times$10$^{3}$ & (8.1 $\pm$ 0.2) & (2.70 $\pm$ 0.03)$\times$10$^{3}$\\   
 & 2 & Undissolved & (2.0 $\pm$ 0.1)$\times$10$^{2}$ & (3.2 $\pm$ 0.2)$\times$10$^{1}$ & (2.40 $\pm$ 0.01)$\times$10$^{3}$ & (1.20 $\pm$ 0.03)$\times$10$^{1}$ & (3.50 $\pm$ 0.06)$\times$10$^{3}$ \\ 
\hline
\end{tabular}}
\caption{Measured fallout rate of Mg, Mn, Fe, Ni, and Zn, in pg$\cdot$day$^{-1}$$\cdot$cm$^{-2}$, in each replicate of the monitored underground locations. For each replicate, information is provided about whether or not visible particulate was observed, and if observed, whether or not it was dissolved or part of it remained undissolved.}
\label{tab:pg}
\end{table}

\begin{table}
\resizebox{\textwidth}{!}{
%\centering
\begin{tabular}{| c | c | c | c |} 
\hline 
 Location & Replicate & Particulate & \multicolumn{1}{|c|}{Accumulation rate [pg$\cdot$day$^{-1}\cdot$cm$^{-2}$]}\\
 \hline
& & & Pb \\
\hline
L6 & 1 & Not observed & (4.9 $\pm$ 1.7)$\times$10$^{-2}$ \\
 & 2 & Undissolved & (2.1 $\pm$ 0.1)$\times$10$^{-1}$ \\
  & 3 & Dissolved & (3.3 $\pm$ 0.5)$\times$10$^{-2}$ \\
  \hline
S1 & 1 & Not observed & (7.6 $\pm$ 0.3)$\times$10$^{1}$ \\  
 & 2 & Not observed & (1.8 $\pm$ 0.1)$\times$10$^{1}$ \\ 
 \hline
S2 & 1 & Not observed & (1.3 $\pm$ 0.1)$\times$10$^{1}$ \\  
 & 2 & Not observed & (1.1 $\pm$ 0.1)$\times$10$^{1}$ \\ 
\hline
\end{tabular}}
\caption{Stable Pb fallout rate, in pg$\cdot$day$\cdot$cm$^{-2}$, measured in L6 during regular activity and in two locations around L6 (S1 and S2) during the construction of a Pb shield. Results from all replicates are reported.}
\label{tab:sensei_Pb}
\end{table}

\begin{table}
\resizebox{\textwidth}{!}{
%\centering
\begin{tabular}{| c | c | c | c c c c c |} 
\hline 
 Location & Replicate & Particulate & \multicolumn{5}{|c|}{Accumulation rate [pg$\cdot$day$^{-1}\cdot$cm$^{-2}$]}\\
 \hline
& & & Mg & Mn & Fe & Ni & Zn\\
\hline
S1 & 1 & Not observed & (8.1 $\pm$ 0.1) & (3.4 $\pm$ 0.2) & (3.6 $\pm$ 0.2) & (1.5 $\pm$ 0.2) & (42.8 $\pm$ 0.2)\\  
 & 2 & Not observed & (9.0 $\pm$ 0.2) & (3.3 $\pm$ 0.1) & (6.7 $\pm$ 1.0) & (1.4 $\pm$ 0.1) & (23.8 $\pm$ 0.3)\\ 
 \hline
S2 & 1 & Not observed & (22.4 $\pm$ 0.3) & (3.5 $\pm$ 0.3) & (2.3 $\pm$ 0.3) & (1.3 $\pm$ 0.2) & (161 $\pm$ 1)\\  
 & 2 & Not observed & (32.5 $\pm$ 0.5) & (3.5 $\pm$ 0.3) & (2.1 $\pm$ 0.3) & (1.4 $\pm$ 0.1) & (249 $\pm$ 1)\\ 
\hline
\end{tabular}}
\caption{Measured fallout rate of Mg, Mn, Fe, Ni, and Zn, in pg$\cdot$day$^{-1}$$\cdot$cm$^{-2}$, in each replicate of the monitored underground location during the lead bricks handling.}
\label{tab:S_others}
\end{table}

\end{document}